\documentclass[11pt]{article}
\usepackage{amsmath,amssymb,epsfig,bm}

\date{\empty}

\begin{document}

\title{\bf Large-scale peculiar velocity fields: Newtonian vs relativistic treatment}

\author{Konstantinos Filippou and Christos G. Tsagas\\ {\small Section of Astrophysics, Astronomy and Mechanics, Department of Physics}\\ {\small Aristotle University of Thessaloniki, Thessaloniki 54124, Greece}}

\maketitle

\begin{abstract}
We employ a perturbative analysis to study the evolution of large-scale peculiar velocity fields within the framework of Newtonian gravity and then compare our results to those of the corresponding relativistic treatment. In so doing, we use the same mathematical formalism and apply the same physical approach. This facilitates a direct and transparent comparison between the two treatments. Our study recovers and extends the familiar Newtonian results on the one hand, while on the other it shows that the Newtonian analysis leads to substantially weaker growth-rates for the peculiar velocity field, compared to the relativistic approach. This implies that, by using Newton's rather than Einstein's theory, one could seriously underestimate the overall kinematic evolution of cosmological peculiar motions. We are also in the position to identify the reason the two theories arrive at such considerably different results and conclusions.
\end{abstract}

\section{Introduction}\label{sI}
Large-scale peculiar velocities are typically treated as a recent addition to the kinematics of the post-recombination universe, triggered by its ever increasing inhomogeneity and anisotropy, both of which reflect the ongoing process of structure formation. The theoretical investigation of the peculiar velocity fields observed in the universe today, namely the study of their evolution and their implications, has a fairly long research history. Nevertheless, essentially all the available cosmological studies are purely Newtonian (e.g.~see~\cite{Pe}-\cite{Pa} and references therein), or they take place within the so-called quasi-Newtonian framework (see~\cite{M}). Also, the available treatments focus on the peculiar velocity itself, rather than the ``irreducible'' kinematic variables of the peculiar motion, namely its expansion/contraction, its shear and its vorticity. A relativistic treatment of large-scale peculiar velocities was used by~\cite{ET}, though in the context of the Zeldovich approximation rather than for studying the evolution of the peculiar velocity field. To the best of our knowledge, the first general relativistic analytical solutions for the full spectrum of the peculiar kinematics were the ones recently given in~\cite{TT}.

The latter work used (relativistic) cosmological perturbation theory to investigate the linear evolution of the peculiar velocity itself, as well as that of its irreducible kinematic components. These are the volume expansion/contraction, the shear distortion and the rotation of the peculiar flow. When compared to the Newtonian and quasi-Newtonian results, those of the relativistic analysis indicated considerably stronger growth-rates for all aspects of the peculiar velocity field, especially on large scales. Motivated by the aforementioned disagreement, we provide here a Newtonian study that will allow for a direct and transparent comparison with the aforementioned quasi-Newtonian and relativistic treatments. This is achieved, by employing the Newtonian version of the 1+3~covariant approach used in~\cite{M} and also in~\cite{TT}.

After a brief introduction to the covariant formalism and its application to the kinematics and the hydrodynamics of a Newtonian fluid, we turn our attention to the study of the peculiar velocity field. In so doing, we consider the Newtonian version of a perturbed Einstein-de Sitter universe, namely we assume a homogeneous and isotropic background cosmology containing a pressureless medium. The latter can be in the form of baryonic or/and low-energy cold dark matter (CDM). Taking the view point of an observer that moves along with the peculiar flow, we obtain the linear evolution equation of the peculiar velocity vector. This in turn provides the propagation formulae of the expansion/contraction, the shear and the vorticity of the peculiar motion. By construction, these relations monitor the linear evolution of the full peculiar kinematics in a perturbed Friedmann-Robertson-Walker (FRW) universe within the framework of the Newtonian theory. Only the homogeneous parts of the aforementioned differential equations accept analytic solutions, which hold on scales where the inhomogeneous component is subdominant. In particular, the peculiar velocity is found to grow with time as $\tilde{v}\propto t^{1/3}$, whereas the peculiar expansion/contraction and the peculiar shear decay as $\tilde{\vartheta}\propto t^{-1/3}$ and $\tilde{\varsigma}\propto t^{-1/3}$ respectively. The rotation (if any) of the peculiar flow, on the other hand, decreases as $\tilde{\varpi}\propto t^{-4/3}$ on all scales. Comparing with the earlier Newtonian and quasi-Newtonian treatments~\cite{Pe}-\cite{M}, confirms the agreement on the evolution rates of the peculiar velocity, of the peculiar expansion/contraction and of the peculiar vorticity. To the best of our knowledge, there are no theoretical Newtonian studies of the peculiar shear to compare with.

Comparing to the results of~\cite{TT}, we find that the Newtonian and the quasi-Newtonian treatments have led to growth rates considerably weaker than those of the general relativistic study. Nevertheless, the disagreement with the quasi-Newtonian analysis is only an apparent one. We show this in \S~\ref{sCRA}, where we provide a relativistic study, this time within the quasi-Newtonian framework. In particular, we find that the aforementioned lack of agreement vanishes when (standard) linear cosmological perturbation theory is employed to express the sources of the peculiar-velocity field and more specifically the 4-acceleration. Moreover, in the process, we also relax the assumptions and thus broaden the range of the analysis given in~\cite{TT}.

The disagreement with the purely Newtonian studies persists, however, due to the different way the two theories address issues as fundamental as the nature of gravity itself. More specifically, in relativity, the energy flux generally contributes to the energy-momentum tensor and therefore to the Einstein field equations. When applied to our case, this means that the energy flux triggered by the peculiar motion of the matter adds to the local gravitational field. In a sense, the drift flow itself gravitates~\cite{TT}. This, purely relativistic effect, modifies the evolution formulae of peculiar-velocity perturbations in ways that the Newtonian theory cannot reproduce. Consequently, the two approaches arrive at dissimilar sets of differential equations for the description of the peculiar kinematics. These, in turn, accept different solutions that lead to unalike results and conclusions. Overall, in contrast to its Newtonian counterpart, the relativistic study appears to favour large-scale peculiar motions faster than it is generally expected.

\section{Newtonian covariant hydrodynamics}\label{sNCHs}
The covariant approach to fluid dynamics originates with the work of~\cite{HS}. The formalism was initially employed in Newtonian studies and later extended to relativistic applications (see~\cite{Eh,El1}, as well as~\cite{TCM,EMM} for recent extensive reviews).

\subsection{Gravitational field and conservation 
laws}\label{ssGFCLs}
In the Newtonian covariant treatment, one introduces spatial coordinates ($x^{\alpha}$, with $\alpha=1,2,3$) and defines the Euclidean metric tensor ($h_{\alpha\beta}$), so that $h_{\alpha}{}^{\alpha}=3$ and $v^2=h_{\alpha\beta}v^{\alpha}v^{\beta}$ for any vector field $v_{\alpha}$.\footnote{Throughout this manuscript Greek indices run from 1 to 3, while Latin ones take values from 0 to 3.} When using a Cartesian reference frame, the above metric coincides with the familiar Kronecker delta (i.e.~$h_{\alpha\beta}= \delta_{\alpha\beta}$). Otherwise, $h_{\alpha\beta}\neq\delta_{\alpha\beta}$ and one needs both $h_{\alpha\beta}$ and $h^{\alpha\beta}$ (with $h_{\alpha\mu}h^{\mu\beta}=\delta_{\alpha}{}^{\beta}$) when raising and lowering tensor indices, to compensate for the ``curvature'' of the coordinate system. In such a case, covariant rather than ordinary partial derivatives should also be used (e.g.~see~\cite{El1,El2}).

We adopt the fluid description, by introducing a vector field ($u_{\alpha}$) that coincides with the velocity of the matter. Then, the time derivative of a general (tensorial) quantity ($T$) is the convective derivative along the motion of the fluid, namely $\dot{T}=\partial_tT+u^{\alpha}\partial_{\alpha}T$. For instance, the (inertial) acceleration of the matter is given by the convective derivative $\dot{u}_{\alpha}=\partial_tu_{\alpha}+ u^{\beta}\partial_{\beta}u_{\alpha}$ of the velocity. Additional kinematic information is encoded in the spatial gradient of the velocity field, which decomposes as~\cite{El1,El2}
\begin{equation}
\partial_{\beta}u_{\alpha}= {1\over3}\,\Theta h_{\alpha\beta}+ \sigma_{\alpha\beta}+ \omega_{\alpha\beta}\,,  \label{pbua}
\end{equation}
with $\Theta=\partial^{\alpha}u_{\alpha}$, $\sigma_{\alpha\beta}= \partial_{\langle\beta}u_{\alpha\rangle}$ and $\omega_{\alpha\beta}= \partial_{[\beta}u_{\alpha]}$.\footnote{Round brackets indicate symmetrisation, square ones antisymmetrisation and angled brackets denote the symmetric traceless part of second-rank tensors. Therefore, $\sigma_{\alpha\beta}= \partial_{(\beta}u_{\alpha)} -(\partial^{\mu}u_{\mu}/3)h_{\alpha\beta}$ by construction.} The former is the volume scalar that describes the expansion/contraction of the fluid, when positive/negative respectively. The symmetric and trace-free shear tensor ($\sigma_{\alpha\beta}$) monitors kinematic anisotropies, while the antisymmetric vorticity tensor ($\omega_{\alpha\beta}$) determines the rotational behaviour of the matter. In cosmological studies the volume scalar is used to define the scale factor ($a=a(t)$) of the universe, by means of $\dot{a}/a=\Theta/3$. Also, starting from the vorticity tensor one obtains the vorticity vector $\omega_{\alpha}= \varepsilon_{\alpha\beta\mu}\omega^{\beta\mu}/2$, which determines the rotational axis. Note that $\varepsilon_{\alpha\beta\mu}$ is the Euclidean Levi-Civita tensor, with $\varepsilon_{\alpha\beta\mu}= \varepsilon_{[\alpha\beta\mu]}$, $\varepsilon_{123}=1$ and $\varepsilon_{\alpha\beta\mu} \varepsilon^{\nu\tau\iota}= 3!\delta_{[\alpha}{}^{\nu} \delta_{\beta}{}^{\tau}\delta_{\mu]}{}^{\iota}$.

The Newtonian gravitational field is monitored by the associated potential ($\Phi$), which is coupled to the matter via the Poisson equation $\partial^2\Phi=\kappa\rho/2$, where $\rho$ is the density of the material component and $\kappa=8\pi G$.\footnote{Applied to a homogeneous system, Poisson's equation leads to an inconsistency that is typically bypassed by appealing to the so-called ``Jeans swindle'' (e.g.~see~\cite{BT}). There is no such problem in relativity.} The spatial gradient of the potential describes the gravitational acceleration, which combines with its inertial counterpart to give
\begin{equation}
A_{\alpha}= \dot{u}_{\alpha}+ \partial_{\alpha}\Phi\,.  \label{Aa}
\end{equation}
The latter expresses the coupled action of inertial and gravitational forces~\cite{El2,ST}. Note that $A_{\alpha}$ corresponds to the relativistic 4-acceleration vector (see \S~\ref{ssRPMs} below), which vanishes when matter moves under inertia or/and gravity alone. On using $A_{\alpha}$, Euler's formula becomes~\cite{El2,ST}
\begin{equation}
\rho A_{\alpha}= -\partial_{\alpha}p- \partial^{\beta}\pi_{\alpha\beta}\,,  \label{Euler}
\end{equation}
with $p$ and $\pi_{\alpha\beta}$ (where $\pi_{\alpha\beta}=\pi_{\beta\alpha}$ and $\pi_{\alpha}{}^{\alpha}=0$) representing the isotropic and the anisotropic pressure (i.e.~the viscosity) of the fluid respectively. Finally, the continuity equation reads
\begin{equation}
\dot{\rho}= -\Theta\rho\,.  \label{cont}
\end{equation}
We also need an equation of state for the pressure. For an ideal medium, the latter typically has the barotropic profile $p=p(\rho)$, with $p=0$ in the case of low energy ``dust''.

\subsection{Kinematics}\label{ssKs}
The kinematic evolution of the matter is described by a set of three propagation and three constraint equations. These monitor the irreducible kinematic variables, namely $\Theta$, $\sigma_{\alpha\beta}$ and $\omega_{\alpha\beta}$ and they all follow from the constraints $\partial_{[t}\partial_{\beta]}u_{\alpha}=0$ and $\partial_{[\mu}\partial_{\beta]}u_{\alpha}=0$. In particular, the trace, the symmetric trace-free and the antisymmetric parts of the former constraint lead to the evolution formulae~\cite{ST}
\begin{equation}
\dot{\Theta}= -{1\over3}\,\Theta^2- {1\over2}\;\kappa\rho- 2\left(\sigma^2-\omega^2\right)+ \partial^{\alpha}A_{\alpha}\,,  \label{Ray}
\end{equation}
\begin{equation}
\dot{\sigma}_{\alpha\beta}= -{2\over3}\,\Theta\sigma_{\alpha\beta}- E_{\alpha\beta}- \sigma_{\mu\langle\alpha}\sigma^{\mu}{}_{\beta\rangle}+ \omega_{\mu\langle\alpha}\omega^{\mu}{}_{\beta\rangle}+ \partial_{\langle\alpha}A_{\beta\rangle}  \label{sigmadot}
\end{equation}
and
\begin{equation}
\dot{\omega}_{\alpha\beta}= -{2\over3}\,\Theta\omega_{\alpha\beta}- 2\sigma_{\mu[\alpha}\omega^{\mu}{}_{\beta]}+ \partial_{[\beta}A_{\alpha]}\,,  \label{omegadot}
\end{equation}
for the volume scalar, the shear and the vorticity tensors respectively. Note the symmetric traceless tensor $E_{ab}= \partial_{\langle\alpha}\partial_{\beta\rangle}\Phi$, which represents tidal forces and closely corresponds to the relativistic electric Weyl tensor. The above propagation equations are supplemented by an equal number of constraints. More specifically, by isolating the trace the symmetric trace-free and the antisymmetric components of $\partial_{[\mu}\partial_{\beta]}u_{\alpha}=0$, one arrives at
\begin{equation}
\partial^{\alpha}\omega_{\alpha}= 0\,, \hspace{20mm} {\rm curl}\sigma_{\alpha\beta}+ \partial_{\langle\alpha}\omega_{\alpha\rangle}=0  \label{constr1}
\end{equation}
and
\begin{equation}
{2\over3}\,\partial_{\alpha}\Theta- \partial^{\beta}\sigma_{\alpha\beta}+ {\rm curl}\omega_{\alpha}= 0\,, \label{constr2}
\end{equation}
respectively~\cite{ST}. Note that ${\rm curl}v_{\alpha}= \varepsilon_{\alpha\beta\mu}\partial^{\beta}v^{\mu}$ for every vector and ${\rm curl}w_{\alpha\beta}= \varepsilon_{\mu\nu\langle\alpha} \partial^{\mu}w_{\beta\rangle}{}^{\nu}$ for every symmetric traceless second-rank tensor.

\section{Newtonian peculiar motions}\label{sNPMs}
In what follows, we will provide a Newtonian covariant treatment of peculiar motions in cosmology. This will allow us to compare with the results of the available Newtonian and quasi-Newtonian treatments, as well as with those of the relativistic analysis.

\subsection{The peculiar kinematics}\label{ssPKs}
Let us consider a pair of relatively moving observers with velocities $u_{\alpha}$ and $\tilde{u}_{\alpha}$ respectively. In Newtonian theory, these two velocity fields are related by the Galilean transformation
\begin{equation}
\tilde{u}_{\alpha}= u_{\alpha}+ \tilde{v}_{\alpha}\,,  \label{Galilean}
\end{equation}
where $\tilde{v}_{\alpha}$ is the peculiar velocity of the $\tilde{u}_{\alpha}$-field relative to the (reference) $u_{\alpha}$-frame.\footnote{Typical Newtonian studies of peculiar motions introduce physical ($r^{\alpha}$) and comoving ($x^{\alpha}$) coordinates, with $r^{\alpha}=ax^{\alpha}$. The time derivative of the latter leads to $v_t=v_H+v_p$, where $v_t=\dot{r}^{\alpha}$, $v_H=Hr^{\alpha}$ and $v_p=a\dot{x}^{\alpha}$ are the total, the Hubble and the peculiar velocities respectively. On an FRW background, the above velocity relation is equivalent to Eq.~(\ref{Galilean}). Also note that the mean peculiar velocity ($\tilde{V}$) is given by the integral $\tilde{V}=(3/4\pi r^3)\int_{x<r}\tilde{v}{\rm d}x^3$, with $\tilde{v}^2= \tilde{v}_{\alpha}\tilde{v}^{\alpha}$ and with $r$ representing the radius of the moving region.} The irreducible kinematics of the $u_{\alpha}$-field are given by decomposition (\ref{pbua}), with an exactly analogous split holding for its tilded counterpart $\tilde{u}_{\alpha}$ (i.e.~$\partial_{\beta}\tilde{u}_{\alpha}= (\tilde{\Theta}/3)h_{\alpha\beta}+\tilde{\sigma}_{\alpha\beta}+ \tilde{\omega}_{\alpha\beta}$). Similarly, the spatial gradient of the peculiar velocity field decomposes as
\begin{equation}
\partial_{\beta}\tilde{v}_{\alpha}= {1\over3}\,\tilde{\vartheta}h_{\alpha\beta}+ \tilde{\varsigma}_{\alpha\beta}+ \tilde{\varpi}_{\alpha\beta}\,,  \label{pbtva}
\end{equation}
with $\tilde{\vartheta}=\partial^{\alpha}\tilde{v}_{\alpha}$, $\tilde{\varsigma}_{\alpha\beta}= \partial_{\langle\beta}\tilde{v}_{\alpha\rangle}$ and $\tilde{\varpi}_{\alpha\beta}= \partial_{[\beta}\tilde{v}_{\alpha]}$ respectively representing the volume scalar, the shear tensor and the vorticity of the peculiar flow. As before (see Eq.~(\ref{pbua}) in \S~\ref{ssGFCLs}), positive/negative values for $\tilde{\vartheta}$ imply that the bulk flow is (locally) expanding/contracting, while nonzero values for $\tilde{\varsigma}_{\alpha\beta}$ and $\tilde{\varpi}_{\alpha\beta}$ indicate local shear deformation and rotation respectively. Starting from transformation (\ref{Galilean}), it is then straightforward to show the following relations
\begin{equation}
\tilde{\Theta}= \Theta+ \tilde{\vartheta}\,, \hspace{10mm} \tilde{\sigma}_{\alpha\beta}= \sigma_{\alpha\beta}+\tilde{\varsigma}_{\alpha\beta} \hspace{10mm} {\rm and} \hspace{10mm} \tilde{\omega}_{\alpha\beta}= \omega_{\alpha\beta}+\tilde{\varpi}_{\alpha\beta}\,,  \label{Nrels}
\end{equation}
between the three kinematic sets. In addition, taking the convective derivative of (\ref{Galilean}), with respect to the tilded frame, and then employing Eq.~(\ref{Galilean}) again, we arrive at the expression
\begin{equation}
\tilde{u}_{\alpha}^{\prime}= \dot{u}_{\alpha}+ \tilde{v}_{\alpha}^{\prime}+ {1\over3}\,\Theta\,\tilde{v}_{\alpha}+ \left(\sigma_{\alpha\beta}+\omega_{\alpha\beta}\right) \tilde{v}^{\beta}\,,  \label{tEuler}
\end{equation}
relating the (inertial) acceleration vectors in the two coordinate systems. Note that primes denote convective derivatives in the tilded frame (i.e.~$\tilde{u}_{\alpha}^{\prime}= \partial_t\tilde{u}_{\alpha}+ \tilde{u}^{\beta}\partial_{\beta}\tilde{u}_{\alpha}$), while overdots indicate convective differentiation in the reference frame (i.e.~$\dot{u}_{\alpha}=\partial_tu_{\alpha}+ u^{\beta}\partial_{\beta}u_{\alpha}$). Following (\ref{tEuler}), the presence of relative motion means that (generally) we cannot set $\dot{u}_{\alpha}$ and $\tilde{u}^{\prime}_{\alpha}$ to zero simultaneously. The same is also true for the rest of the kinematic variables (see Eqs.~(\ref{Nrels}a)-(\ref{Nrels}c) above). Finally, we should point out that, in contrast to the relativistic treatment (see~\S~\ref{ssRPMs} later), the matter variables remain unchanged when transforming from one coordinate system to the other.

\subsection{Linear sources of peculiar velocities}\label{ssLSPVs}
So far our analysis has been nonlinear. Let us now consider a perturbed almost-FRW Newtonian universe that contains an ideal pressureless fluid (baryonic or/and CDM -- with $p=0= \pi_{\alpha\beta}$) and treat the peculiar velocity field as a perturbation on the aforementioned background. Then, by identifying the $u_{\alpha}$-field with the rest-frame of the Cosmic Microwave Background (CMB), namely with the coordinate system of the smooth (isotropic) Hubble flow, we may set $\dot{u}_{\alpha}=0= \sigma_{\alpha\beta}=\omega_{\alpha\beta}$.\footnote{By construction, the CMB frame is the coordinate system where the CMB dipole vanishes (e.g.~\cite{TCM,EMM}). Put another way, the CMB frame is the rest-system of the (fictitious) idealised observers that follow the smooth Hubble expansion. Real observers, living in typical galaxies like our Milky Way, have nonzero peculiar velocities relative to the Hubble frame. Clearly, in the unperturbed background, these two coordinate systems coincide.} On these grounds, expression (\ref{tEuler}) linearises to
\begin{equation}
\tilde{v}_{\alpha}^{\prime}= -H\tilde{v}_{\alpha}+ \tilde{u}_{\alpha}^{\prime}\,,  \label{ltEuler1}
\end{equation}
where $H=\dot{a}/a$ is the background Hubble parameter. According to the above, linear peculiar velocities are triggered by the acceleration, which in the absence of pressure is given by $\tilde{u}_{\alpha}^{\prime}= -\partial_{\alpha}\Phi$ (see Eqs.~(\ref{Aa}) and (\ref{Euler}) in \S~\ref{ssGFCLs} earlier). As a result, (\ref{ltEuler1}) recasts into
\begin{equation}
\tilde{v}_{\alpha}^{\prime}= -H\tilde{v}_{\alpha}- \partial_{\alpha}\Phi\,.  \label{ltv'}
\end{equation}
This is the linear propagation equation of peculiar velocities in a perturbed, Newtonian Einstein-de Sitter universe, relative to the $\tilde{u}_{\alpha}$-frame. Relation (\ref{ltv'}) is formally identical to the one obtained in the quasi-Newtonian treatments of~\cite{M}, as well as to those given in typical Newtonian studies, provided the latter equations are written in physical (rather than comoving) coordinates (e.g.~see~\cite{Pe,NDBB}).

The evolution formulae of the irreducible peculiar kinematic variables, namely of the expansion/contraction, the shear and the vorticity, follow from the spatial gradient of Eq.~(\ref{ltv'}). Indeed, keeping up to first-order terms and recalling that $\partial_{\beta}u_{\alpha}=(\Theta/3)h_{\alpha\beta}= Hh_{\alpha\beta}$ to zero perturbative order (see Eq.~(\ref{pbua}) in \S~\ref{ssGFCLs}), we obtain
\begin{equation}
\left(\partial_{\beta}\tilde{v}_{\alpha}\right)^{\prime}= -2H\partial_{\beta}\tilde{v}_{\alpha}- \partial_{\beta}\partial_{\alpha}\Phi\,.  \label{lpbtva'}
\end{equation}
Isolating the trace, the symmetric trace-free and the antisymmetric parts of the above, leads to the linear evolution formula of the peculiar volume scalar
\begin{equation}
\tilde{\vartheta}^{\prime}= -2H\tilde{\vartheta}- \partial^2\Phi\,,  \label{ltvtheta'}
\end{equation}
of the peculiar shear
\begin{equation}
\tilde{\varsigma}^{\prime}_{\alpha\beta}= -2H\tilde{\varsigma}_{\alpha\beta}- \partial_{\langle\beta}\partial_{\alpha\rangle}\Phi  \label{ltvsigma'}
\end{equation}
and of the peculiar vorticity
\begin{equation}
\tilde{\varpi}^{\prime}_{\alpha\beta}= -2H\tilde{\varpi}_{\alpha\beta}\,,  \label{ltvpi'}
\end{equation}
respectively. Expressions (\ref{ltv'}), (\ref{ltvtheta'}) and (\ref{ltvsigma'}) reveal that, in the absence of pressure, the gravitational forces are the sole sources of peculiar velocity perturbations. More specifically the presence of matter perturbations distorts the volume expansion/contraction of the peculiar flow, while tidal forces do the same for the peculiar shear (see Eqs.~(\ref{ltvtheta'}) and (\ref{ltvsigma'}) respectively). On the other hand, since $\partial_{[\beta}\partial_{\alpha]}\Phi=0$, expression (\ref{ltvpi'}) ensures that there are no linear sources of peculiar vorticity. Then, the last differential equation solves immediately to ensure that (after equipartition when $H=2/3t$) the Newtonian peculiar vorticity depletes as
\begin{equation}
\tilde{\varpi}= \mathcal{C}_1t^{-4/3}= \mathcal{C}_2a^{-2}\,,  \label{ltvpi}
\end{equation}
on all scales. This result, which is in agreement with the one quoted in~\cite{Pe}, also means that $\tilde{\varpi}$ decays as the vorticity proper in perturbed pressure-free FRW universes (e.g.~see~\cite{H,EBH}). As a result, the relative strength of the peculiar vorticity drops as $\tilde{\varpi}/H\propto t^{-1/3}\propto a^{-1/2}$. Note that the absence of source terms on the right-hand side of (\ref{ltvpi'}) marks a distinctive difference between the Newtonian and the relativistic treatment of (peculiar) vorticity. This, in turn, highlights the unconventional behaviour of rotating spacetimes in the geometrical framework of Einstein's gravitational theory (see \S~4.4 in~\cite{TT}).

\section{The linear peculiar velocity field}\label{LPVF}
With the exception of the peculiar vorticity, the linear evolution of which has already been determined, the rest of the peculiar kinematics require further study. We will do so next, by taking higher-order derivatives of the associated variables.

\subsection{Linear evolution of the peculiar velocity}\label{ssLEPV}
Taking the convective derivative of (\ref{ltv'}), recalling that $\dot{H}=-3H^2/2$ and that $\partial_{\beta}u_{\alpha}= Hh_{\alpha\beta}$ in the background, while keeping up to linear-order terms, we arrive at
\begin{equation}
\tilde{v}_{\alpha}^{\prime\prime}= -2H\tilde{v}_{\alpha}^{\prime}+ {1\over2}\,H^2\tilde{v}_{\alpha}- \partial_{\alpha}\Phi^{\prime}\,.  \label{ltv''1}
\end{equation}
The above differential formula does not accept analytic solutions unless we isolate its homogeneous part, which is like assuming that  $\partial_{\alpha}\Phi^{\prime}\simeq0$ (while keeping in mind that $\partial_{\alpha}\Phi\neq0$).\footnote{It should be made clear that the term homogeneous/inhomogeneous refers to the type of the differential equation and not to the homogeneity/inhomogenetiy of the space.} In practice, this means that the analytic solution given below applies on relatively large scales, where there are no temporal changes of the gravitational field, or (if they exist) they vary slowly in space. On these wavelengths, after equipartition (when $a\propto t^{2/3}$ and $H=2/3t$), Eq.~(\ref{ltv''1}) reduces to
\begin{equation}
9t^2\,{{\rm d}^2\tilde{v}\over{\rm d}t^2}+ 12t\,{{\rm d}\tilde{v}\over{\rm d}t}- 2\tilde{v}= 0  \label{ltv''2}
\end{equation}
and accepts the power-law solution
\begin{equation}
\tilde{v}= \mathcal{C}_1t^{1/3}+ \mathcal{C}_2t^{-2/3}= \mathcal{C}_3a^{1/2}+ \mathcal{C}_4a^{-1}\,.  \label{ltv}
\end{equation}
Therefore, within the framework of Newtonian gravity and the limits of our approximation, peculiar velocities grow as $\tilde{v}\propto t^{1/3}\propto a^{1/2}$. This means that the dimensionless ratio $\tilde{v}/v_H$, where $v_H=\lambda H\propto t^{-1/3}\propto a^{-1/2}$ is the Hubble velocity on a scale $\lambda$, increases as $\tilde{v}/v_H\propto t^{2/3}\propto a$ after decoupling. These results are in full agreement with the (also Newtonian) analysis of~\cite{Pe}, as well as with the quasi-Newtonian studies of~\cite{M}, but not with the relativistic treatment of~\cite{TT} (see also \S~\ref{ssLRPVs} here).

\subsection{Linear evolution of the peculiar volume scalar and 
shear}\label{ssLEPVSS}
Proceeding in an exactly analogous manner, one obtains the differential formulae monitoring the irreducible peculiar kinematics. In particular, the convective derivative of Eq.~(\ref{lpbtva'}) gives
\begin{equation}
\left(\partial_{\beta}\tilde{v}_{\alpha}\right)^{\prime\prime}= -4H\left(\partial_{\beta}\tilde{v}_{\alpha}\right)^{\prime}- H^2\partial_{\beta}\tilde{v}_{\alpha}- \partial_{\beta}\partial_{\alpha}\Phi^{\prime}\,,  \label{lpbtva''}
\end{equation}
to first approximation. Then, taking the trace and the symmetric traceless components of the above, we arrive at
\begin{equation}
\tilde{\vartheta}^{\prime\prime}= -4H\tilde{\vartheta}^{\prime}- H^2\tilde{\vartheta}- \partial^2\Phi^{\prime}  \label{tvtheta''1}
\end{equation}
and
\begin{equation}
\tilde{\varsigma}_{\alpha\beta}^{\prime\prime}= -4H\tilde{\varsigma}_{\alpha\beta}^{\prime}- H^2\tilde{\varsigma}_{\alpha\beta}- \partial_{\langle\beta}\partial_{\alpha\rangle}\Phi^{\prime}\,,  \label{tvsigma''1}
\end{equation}
respectively. As before, let us assume that the $\Phi^{\prime}$-field is nearly homogeneously distributed. We may then set $\partial^2\Phi^{\prime}\simeq0$ and recast (\ref{tvtheta''1}) into
\begin{equation}
9t^2\,{{\rm d}^2\tilde{\vartheta}\over{\rm d}t^2}+ 24t\,{{\rm d}\tilde{\vartheta}\over{\rm d}t}+ 4\tilde{\vartheta}= 0\,.  \label{tvtheta''2}
\end{equation}
After equipartition, when $a\propto t^{2/3}$ and $H=2/3t$, the latter accepts the power-law solution
\begin{equation}
\tilde{\vartheta}= \mathcal{C}_1t^{-1/3}+ \mathcal{C}_2t^{-4/3}= \mathcal{C}_3a^{-1/2}+ \mathcal{C}_4a^{-2}\,.  \label{ltvtheta}
\end{equation}
Under analogous conditions, namely for $\partial_{\langle b}\partial_{a\rangle}\Phi^{\prime}\simeq0$, expression (\ref{tvsigma''1}) solves to give
\begin{equation}
\tilde{\varsigma}= \mathcal{C}_1t^{-1/3}+ \mathcal{C}_2t^{-4/3}= \mathcal{C}_3a^{-1/2}+ \mathcal{C}_4a^{-2}\,.  \label{ltvsigma}
\end{equation}
Consequently, on scales where the spatial gradients of $\Phi^{\prime}$ are negligible, both the peculiar volume scalar and the peculiar shear decrease as $\tilde{\vartheta}$, $\tilde{\varsigma}\propto t^{-1/3}\propto a^{-1/2}$. Then, $\tilde{\vartheta}/H$, $\tilde{\varsigma}/H\propto t^{2/3}\propto a$ after matter-radiation equality. Overall, as the universe advances into its post-recombination epoch, the linear peculiar kinematics increase (relative to the background expansion), with the exception of the peculiar vorticity (see solution (\ref{ltvpi}) in \S~\ref{ssLSPVs} earlier). Note that our results for $\tilde{\vartheta}$ and $\varpi$ are in agreement with previous analogous treatments (e.g.~see~\cite{Pe,Pa}). To the best of our knowledge, however, there are no analytical Newtonian studies of the peculiar shear.

Next, we will compare the results of our Newtonian treatment with those obtained in~\cite{TT}. The latter study, where we refer the reader for further details and discussion, employs the relativistic version of our 1+3 formalism and also adopts the same physical approach and makes the same approximations.

\section{Comparison to the relativistic analysis}\label{sCRA}
In relativity space and time are no longer absolute and separate entities. Also, there is no gravitational potential,  but spacetime curvature, while Poisson's formula has been replaced by the Einstein field equations. Next, we will demonstrate how these fundamental differences affect the study of cosmological peculiar-velocity fields.

\subsection{Relativistic peculiar motions}\label{ssRPMs}
Assuming a pair of observers moving with respect to each other, the relativistic analogue of the Galilean transformation seen in Eq.~(\ref{Galilean}) is the familiar Lorentz boost
\begin{equation}
\tilde{u}_a= \tilde{\gamma}\left(u_a+\tilde{v}_a\right)\,,  \label{Lorentz}
\end{equation}
where $\tilde{u}_a$ and $u_a$ are the (timelike) 4-velocity vectors of the aforementioned observers and $\tilde{v}_a$ is the (spacelike) peculiar velocity of the former relative to the latter (e.g.~see~\cite{ET,TCM,EMM}). Note that $\tilde{u}_a\tilde{u}^a=-1=u_au^a$, $u_a\tilde{v}^a=0$ and $\tilde{\gamma}=1/\sqrt{1-\tilde{v}^2}$ by construction. Also, $\cosh\beta=-\tilde{u}_au^a=\tilde{\gamma}$ defines the hyperbolic ``tilt'' angle ($\beta$) between the two 4-velocity fieds (see Fig.~\ref{fig:bflow} in \S~\ref{sssR4-A} below). When dealing with non-relativistic peculiar motions, we have $\tilde{v}^2\ll1$, $\tilde{\gamma}\simeq1$ and the above reduces to $\tilde{u}_a\simeq u_a+\tilde{v}_a$. This relation should not be confused with the Galilean transformation of Eq.~(\ref{Galilean}), despite their close resemblance, since both $\tilde{u}_a$ and $u_a$ remain timelike 4-vectors.

Each 4-velocity field defines a temporal direction and introduces a spatial section orthogonal to it. The symmetric tensors $h_{ab}=g_{ab}+u_au_b$ and $\tilde{h}_{ab}=g_{ab}+ \tilde{u}_a\tilde{u}_b$ (with $h_{ab}u^b=0= \tilde{h}_{ab}\tilde{u}^b$ and $h_a{}^a=\tilde{h}_a{}^a=3$) project into these 3-dimensional hypersurfaces. Also, the operators $\;{}^{\cdot}=u^a\nabla_a$ and ${\rm D}_a= h_a{}^b\nabla_b$ respectively define temporal and spatial differentiation in the $u_a$-frame. Similarly, the set $\;{}^{\prime}= \tilde{u}^a\nabla_a$ and $\tilde{\rm D}_a= \tilde{h}_a{}^b\nabla_b$
denotes time and 3-space derivatives in the ``tilted'' frame. Then, the gradients of the two 4-velocity vectors decompose as~\cite{ET,TCM,EMM}
\begin{equation}
\nabla_bu_a= {1\over3}\,\Theta h_{ab}+ \sigma_{ab}+ \omega_{ab}- A_au_b \hspace{10mm} {\rm and} \hspace{10mm} \nabla_b\tilde{u}_a= {1\over3}\,\tilde{\Theta}\tilde{h}_{ab}+ \tilde{\sigma}_{ab}+ \tilde{\omega}_{ab}- \tilde{A}_a\tilde{u}_b\,,  \label{Nbua}
\end{equation}
with $\Theta={\rm D}^au_a$, $\sigma_{ab}={\rm D}_{\langle b}u_{a\rangle}$, $\omega_{ab}={\rm D}_{[b}u_{a]}$ and $A_a=\dot{u}_a$ being the volume scalar, the shear tensor, the vorticity tensor and the 4-acceleration vector of the relativistic analysis. Exactly analogous relations define the corresponding variables in the tilted frame. Relative to the same coordinate system, the gradient of the peculiar velocity field splits as~\cite{ET,TT}
\begin{equation}
\tilde{\rm D}_b\tilde{v}_a= {1\over3}\,\tilde{\vartheta}\tilde{h}_{ab}+ \tilde{\varsigma}_{ab}+ \tilde{\varpi}_{ab}\,,  \label{Dbtva}
\end{equation}
where $\tilde{\vartheta}=\tilde{\rm D}^a\tilde{v}_a$, $\tilde{\varsigma}_{ab}=\tilde{\rm D}_{\langle b}\tilde{v}_{a\rangle}$ and $\tilde{\varpi}_{ab}=\tilde{\rm D}_{[b}\tilde{v}_{a]}$ are the relativistic counterparts of the kinematic quantities defined in \S~\ref{ssPKs} earlier. When dealing with small peculiar velocities (so that $\tilde{v}^2\ll1$) in a perturbed FRW model, the linear relations between the above three sets read~\cite{M}
\begin{eqnarray}
\tilde{\Theta}= \Theta+ \tilde{\vartheta}\,, &\hspace{10mm}& \tilde{\sigma}_{ab}= \sigma_{ab}+ \tilde{\varsigma}_{ab}\,,  \label{lRGrels1}\\ \tilde{\omega}_{ab}= \omega_{ab}+ \tilde{\varpi}_{ab} \hspace{10mm} &{\rm and}& \hspace{10mm} \tilde{A}_a= A_a+ \tilde{v}_a^{\prime}+ {1\over3}\,\Theta\tilde{v}_a\,.  \label{lGRrels2}
\end{eqnarray}
These coincide with their Newtonian analogues (compare to Eqs.~(\ref{Nrels}) and (\ref{tEuler}) in \S~\ref{ssPKs}). Note, in particular, that (\ref{lGRrels2}b) reduces to the linearised version of (\ref{tEuler}) when the 4-acceleration vectors ($A_a$ and $\tilde{A}_a$) are replaced by their inertial counterparts ($\dot{u}_{\alpha}$ and $\tilde{u}_{\alpha}^{\prime}$ respectively). Also, as in the Newtonian case, expression (\ref{lGRrels2}b) ensures that we should not set both 4-acceleration vectors to zero simultaneously, when peculiar motions are present.\footnote{Setting both $A_a$ and $\tilde{A}_a$ to zero imposes a strict constraint on the peculiar velocity field, demanding that $\tilde{v}_a^{\prime}=-(\Theta/3)\tilde{v}_a$. The latter implies that $\tilde{v}_a\propto a^{-1}$ on all scales. Given that peculiar velocities start very weak at recombination, such a result is incompatible with the observations and with the widespread presence of large-scale bulk flows in the universe (see also \S~3.2 in~\cite{TT} for additional discussion).} The same also holds for the rest of the kinematic variables.

In Newtonian theory, relative motion does not ``alter'' the nature of the matter fields involved. This is no longer true in relativity, where we have the following linear relations~\cite{M}
\begin{equation}
\tilde{\rho}= \rho\,, \hspace{10mm} \tilde{p}= p\,, \hspace{10mm} \tilde{q}_a= q_a- (\rho+p)\tilde{v}_a \hspace{5mm} {\rm and} \hspace{5mm} \tilde{\pi}_{ab}= \pi_{ab}\,,  \label{lGRrels3}
\end{equation}
between the energy density ($\rho$), the isotropic pressure ($p$), the energy flux ($q_a$) and the viscosity ($\pi_{ab}$) of the matter. According to (\ref{lGRrels3}b) and (\ref{lGRrels3}d), the matter pressure can vanish in both frames simultaneously. This is not the case for the energy flux, however, since $q_a=0$ implies $\tilde{q}_a=-(\rho+p)\tilde{v}_a\neq0$ (see Eq.~(\ref{lGRrels3}c) above). In relativity, the energy flux contributes to the stress-energy tensor and therefore it gravitates (e..g.~see~\cite{TCM,EMM}). Therefore, when dealing with peculiar motions, there is a net flux input to the local gravitational field explicitly due to the matter flow. Put another way, the bulk flow itself gravitates~\cite{TT}. As we will outline next, this effect feeds into the equations of relativistic cosmological perturbation theory and eventually into the formulae governing the evolution of the peculiar velocity field.

\subsection{Linear relativistic peculiar velocities}\label{ssLRPVs}
As in the Newtonian study (see \S~\ref{ssLSPVs} earlier), let us consider a perturbed, almost-FRW universe filled with a pressureless fluid (baryonic or/and CDM).\footnote{The assumption of a Friedmann background, where all perturbations (including the peculiar velocity) vanish, ensures the gauge invariance of the study~\cite{SW}} We will also allow for two families of observers moving along the matter and along an idealised reference frame.

\subsubsection{The role of the 4-acceleration}\label{sssR4-A}
\begin{figure}[tbp]
\centering \vspace{6cm} \includegraphics{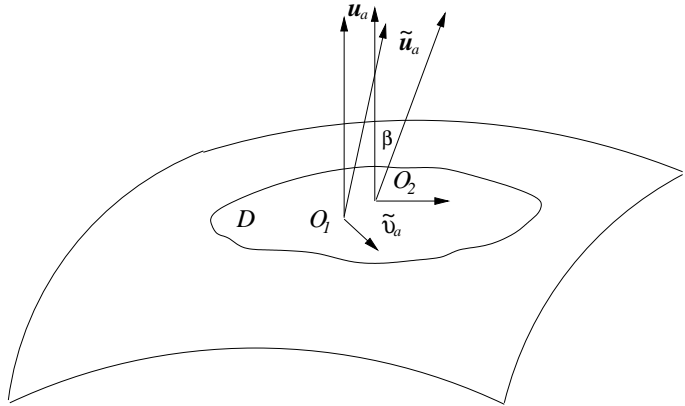} \caption{At every point ($O_1$, $O_2$) in 3-space we define two 4-velocity fields. The matter follows the $\tilde{u}_a$-field, which forms a (hyperbolic) tilt angle ($\beta$) with the quasi-Newtonian frame ($u_a$). Assuming that the matter has (non-relativistic) peculiar velocity $\tilde{v}_a$ with respect to the $u_a$-frame, the latter moves with peculiar velocity $v_a=-\tilde{v}_a$ relative to the matter.}  \label{fig:bflow}
\end{figure}

In the study of~\cite{TT}, the authors took the viewpoint of the real observers, namely those living in a typical galaxy (like our Milky Way) and following the motion of the matter. Here, instead, we will analyse the linear evolution of the peculiar velocity field in the quasi-Newtonian frame used in~\cite{M}. Assuming that $\tilde{v}_a$ is the peculiar velocity of the matter with respect to the quasi-Newtonian coordinate system, the latter moves relative to the matter with velocity $v_a=-\tilde{v}_a$ (see Fig.~\ref{fig:bflow}). Following~\cite{M}, we also set $\sigma_{ab}=0=\omega_{ab}$ and $\tilde{A}_a=0= \tilde{q}_a$.\footnote{Treating the $u_a$-field as both shear-free and irrotational at the linear level is not essential for this study, as well as for that of~\cite{TT}, since neither of these variables is involved in the subsequent calculations. One may therefore relax these kinematic assumptions and in so doing broaden the range of both studies. Nevertheless, setting $\sigma_{ab}=0= \omega_{ab}$ facilitates the connection with the earlier quasi-Newtonian studies (see also footnotes~12, 13 below). Note that the vanishing of the 4-acceleration in the tilted frame of the matter follows from the absence of pressure. On the other hand, setting $\tilde{q}_a=0$ implies aligning the $\tilde{u}_a$-field with the energy frame, where the energy flux vanishes, though the particle flux is generally nonzero.} Then, equations (\ref{lGRrels2}b) and (\ref{lGRrels3}c) reduce to
\begin{equation}
\dot{v}_a= -Hv_a+ A_a \hspace{10mm} {\rm and} \hspace{10mm} q_a=- \rho v_a\,,  \label{lGRv'}
\end{equation}
respectively~\cite{M}. Recall that $v_a=-\tilde{v}_a$, which ensures that $\tilde{v}_a^{\prime}=-\dot{v}_a$ to  first order.\footnote{In~\cite{M}, $v_a$ is the peculiar velocity of the matter relative to the quasi-Newtonian frame. This means that one can recover Eqs.~(\ref{lGRv'}a) and (\ref{lGRv'}b) from the expressions given in~\cite{M} by simply replacing $v_a$ with $-v_a$ in those relations.}

Expressions (\ref{ltEuler1}) and (\ref{lGRv'}a) reveal the decisive role of the acceleration for the subsequent evolution of the peculiar velocity field. The Newtonian acceleration is given by the spatial gradient of the gravitational potential (see Eq.~(\ref{ltv'}) in \S~\ref{ssLSPVs}). In relativity, on the other hand, the 4-acceleration follows from the momentum-density conservation law and subsequently it emerges, through cosmological perturbation theory, in the evolution formula of the density gradients. In particular, combining (\ref{lGRv'}a) and (\ref{lGRv'}b), we obtain $A_a=-(\dot{q}_a+4Hq_a)/\rho$, which substituted into the linear propagation equation of density inhomogeneities (see Eq.~(2.3.1) in~\cite{TCM} and/or Eq.~(10.101) in~\cite{EMM}) leads to~\cite{TK}
\begin{equation}
A_a= {1\over3H}\,{\rm D}_a\vartheta- {1\over3aH}\left(\dot{\Delta}_a +{\mathcal{Z}}_a\right)\,,  \label{lGRAa}
\end{equation}
with $\vartheta={\rm D}^av_a=-{\rm D}^aq_a/\rho=-\tilde{\vartheta}$ at the linear level. Also, the variables $\Delta_a=a{\rm D}_a\rho/\rho$ and $\mathcal{Z}_a=a{\rm D}_a\Theta$ respectively describe spatial gradients in the matter energy density and in the volume expansion of the universe (e.g.~see~\cite{TCM,EMM}). Expression (\ref{lGRAa}), which provides an additional independent relation for the 4-acceleration and also carries the aforementioned contribution of the peculiar energy-flux to the local gravitational field, is the reason for the differences between the Newtonian and the relativistic results presented here.\footnote{The Newtonian version of Eq.~(\ref{lGRAa}) reads $\dot{\Delta}_{\alpha}= -\mathcal{Z}_{\alpha}$, with $\Delta_{\alpha}= (a/\rho)\partial_{\alpha}\rho$ and $\mathcal{Z}_{\alpha}= a\partial_{\alpha}\tilde{\Theta}$~\cite{El2}. The absence of an acceleration term in the evolution formula of $\Delta_{\alpha}$, which reflects the fact that there is no energy-flux contribution to the gravitational field in Newton's theory, explains why expression (\ref{lGRAa}) has no close analogue in Newtonian gravity and why the latter cannot reproduce the relativistic results.}

Taking the time derivative of (\ref{lGRAa}), using the background relations $H^2=\kappa\rho/3$ and $\dot{H}=-\kappa\rho/2$ and keeping up to first-order terms, one arrives at
\begin{equation}
\dot{A}_a= {1\over2}\,HA_a+ {1\over3H}\,{\rm  D}_a\dot{\vartheta}- {1\over3aH}\,\left(\ddot{\Delta}_a+\dot{\mathcal{Z}}_a\right)\,.  \label{lGRdotAa}
\end{equation}
Note that in deriving the above we have assumed a spatially flat FRW background and used the linear commutation law $({\rm D}_a\vartheta)^{\cdot}={\rm D}_a\dot{\vartheta}-H{\rm D}_a\vartheta$.\footnote{In the quasi-Newtonian studies, the 4-acceleration is given by the gradient of an effective gravitational potential (i.e.~$A_a={\rm D}_a\varphi$, with $\varphi$ representing the potential). This is possible due to the irrotational and shear-free nature of the associated reference frame. Also, the  propagation formula of $A_a$ is obtained after introducing the ansatz $\dot{\varphi}=-\Theta/3$ for the evolution of the potential. Then,  instead of Eq.~(\ref{lGRdotAa}), one arrives at the expression $\dot{A}_a=-2HA_a- (\kappa\rho/2)v_a$~\cite{M}. Here, we do not need to employ an effective potential, which explains why the irrotational and shear-free assumption for the $u_a$-field is not essential for our purposes (though it helps to associate with the quasi-Newtonian treatments -- see also footnote~9 earlier and footnote~13 next).}

\subsubsection{The evolution formulae}\label{sssEF}
The system of (\ref{lGRv'}a) and (\ref{lGRdotAa}) governs the relativistic evolution of linear peculiar velocities in a perturbed Friedmann universe during its post-recombination epoch. More specifically, substituting (\ref{lGRdotAa}) into the time derivative of (\ref{lGRv'}a) leads to
\begin{equation}
\ddot{v}_a= -{1\over2}\,H\dot{v}_a+ 2H^2v_a+ {1\over3H}\,{\rm D}_a\dot{\vartheta}- {1\over3aH}\, \left(\ddot{\Delta}_a +\dot{\mathcal{Z}}_a\right)\,.  \label{lGRv''}
\end{equation}
Differentiating the above in 3-space provides the linear propagation  to equation
\begin{equation}
\left({\rm D}_bv_a\right)^{\cdot\cdot}= -{5\over2}\,H\left({\rm D}_bv_a\right)^{\cdot}+ 2H^2{\rm D}_bv_a+ {1\over3H}\,{\rm D}_b{\rm D}_a\dot{\vartheta}- {1\over3a^2H}\, \left(\ddot{\Delta}_{ab} +\dot{\mathcal{Z}}_{ab}\right)\,,  \label{lGRDv''}
\end{equation}
for the peculiar velocity gradient (with $\Delta_{ab}=a{\rm D}_b\Delta_a$ and $\mathcal{Z}_{ab}=a{\rm D}_b\mathcal{Z}_a$ by definition (e.g.~see~\cite{TCM,EMM}). Expressions (\ref{lGRv''}) and (\ref{lGRDv''}) are the general relativistic analogues of the Newtonian relations (\ref{ltv''1}) and (\ref{lpbtva''}) derived earlier here. The differences between the two sets of equations are clear and the underlying reason is that Eqs.~(\ref{lGRv''}), (\ref{lGRDv''}) account for the (purely relativistic) energy-flux contribution of the peculiar flow to the local gravitational field (see also \S~3,~4 in~\cite{TT} for further discussion).

Finally, as with Eq.~(\ref{lpbtva''}), the trace, the symmetric traceless and the antisymmetric parts of (\ref{lGRDv''}) monitor the evolution of the peculiar volume scalar ($\tilde{\vartheta}$), of the peculiar shear ($\tilde{\varsigma}_{ab}$) and of the peculiar vorticity ($\tilde{\varpi}_{ab}$) respectively. In particular, we find that
\begin{equation}
\ddot{\vartheta}= -{5H\over2}\,\dot{\vartheta}+ 2H^2\vartheta+ {1\over3H}\,{\rm D}^2\dot{\vartheta}- {1\over3a^2H}\left(\ddot{\Delta}+\dot{\mathcal{Z}}\right)\,,  \label{lGRvtheta''}
\end{equation}
\begin{equation}
\ddot{\varsigma}_{ab}= -{5\over2}\,H\dot{\varsigma}_{ab}+ 2H^2\varsigma_{ab}+ {1\over3H}\,{\rm D}_{\langle b}{\rm D}_{a\rangle}\dot{\vartheta}- {1\over3a^2H}\left(\ddot{\Delta}_{\langle ab\rangle} +\dot{\mathcal{Z}}_{\langle ab\rangle}\right)  \label{lGRvsigma''}
\end{equation}
and
\begin{equation}
\ddot{\varpi}_{ab}= -{5\over2}\,H\dot{\varpi}_{ab}+ 2H^2\varpi_{ab}- {1\over3a^2H}\left(\ddot{\Delta}_{[ab]} +\dot{\mathcal{Z}}_{[ab]}\right)\,,  \label{lGRvpi''}
\end{equation}
at the linear perturbative level. Note that $\Delta=\Delta_a{}^a$, $\mathcal{Z}=\mathcal{Z}_a{}^a$ are the traces of $\Delta_{ab}$ and $\mathcal{Z}_{ab}$, while $\varsigma_{ab}={\rm D}_{\langle b}v_{a\rangle}$ and $\varpi_{ab}={\rm D}_{[b}v_{a]}$ by definition (with $\varsigma_{ab}= -\tilde{\varsigma}_{ab}$ and $\varpi_{ab}=-\tilde{\varpi}_{ab}$ to first approximation).\footnote{Assuming that the $u_a$-field is both irrotational and shear-free, the peculiar vorticity vanishes at the linear perturbative level~\cite{M}. Relaxing the nonzero shear constraint, while keeping the rotation to zero, ensures that the antisymmetric components of both $\Delta_{ab}$ and $\mathcal{Z}_{ab}$ vanish to first approximation (given that $\Delta_{[ab]}=-3a^2H\omega_{ab}$ and $\mathcal{Z}_{[ab]}=3a^2\dot{H}\omega_{ab}$ to linear order~\cite{EBH}). In the absence of such vortex-like distortions, the linear evolution of the peculiar vorticity is governed by the homogeneous differential equation $\dot{\varpi}_{ab}=-2H\varpi_{ab}$. After equipartition, the latter accepts the solution $\varpi\propto t^{-4/3}\propto a^{-2}$, just like its Newtonian counterpart (see Eq.~(\ref{ltvpi'}) in \S~\ref{ssLSPVs} earlier).}

\subsubsection{The solutions}\label{sssSs}
Expressions (\ref{lGRv''}) and (\ref{lGRvtheta''})-(\ref{lGRvpi''}) are a set of inhomogeneous differential equations, of which only the homogeneous components accept analytic solutions. Recall that the same was also true for the Newtonian differential formulae (\ref{ltv''1}) and (\ref{lpbtva''}), obtained earlier in \S~\ref{ssLEPV} and \S~\ref{ssLEPVSS} respectively.\footnote{As stated in footnote~6, the term homogeneous/inhomogeneous refers to the type of the differential equation and not to the homogeneity/inhomogenetiy of the host spacetime, or of the perturbations. Note that isolating and solving the homogeneous components of the relativistic differential equations ensures that we proceed on equal footing with the Newtonian study and therefore facilitates the direct comparison of the two treatments. Physically speaking, neglecting the inhomogeneous parts of (\ref{lGRv''}) and (\ref{lGRDv''}) implies taking the long-wavelength limit of their solutions (see also \S~4.2 in~\cite{TT}).} The only exception was the Newtonian peculiar vorticity (see Eq.~(\ref{ltvpi'}) in \S~\ref{ssLSPVs}), which obeyed a homogeneous differential equation.

Isolating the homogeneous component of (\ref{lGRv''}) and keeping in mind that $H=2/3t$ after equipartition, we obtain
\begin{equation}
9t^2\ddot{v}_a+ 3t\dot{v}_a- 8v_a=0\,,  \label{lhGRtv''}
\end{equation}
on all scales where the inhomogeneous component is negligible. The above accepts the analytic power-law solution
\begin{equation}
v= \mathcal{C}_1t^{4/3}+ \mathcal{C}_2t^{-2/3}= \mathcal{C}_3a^2+ \mathcal{C}_4 a^{-1}\,,  \label{lRtv}
\end{equation}
which implies growth with $\tilde{v}\propto t^{4/3}\propto a^2$ for the peculiar velocity field. This is significantly stronger than the Newtonian and the quasi-Newtonian growth rates of $\tilde{v}\propto t^{1/3}\propto a^{1/2}$ (see solution (\ref{ltv}) in \S~\ref{ssLEPV} and references therein), implying a growth-rate of $v/v_H\propto t^{5/3}\propto a^{5/2}$ relative to the background Hubble velocity.

Similarly, by removing the inhomogeneous part of Eq.~(\ref{lGRvtheta''}) and then harmonically decomposing the remaining formula, we arrive at
\begin{equation}
9t^2\ddot{\vartheta}_{(n)}+ 15t\left[1+{2\over15}\left({\lambda_H\over\lambda_n}\right)^2\right] \dot{\vartheta}_{(n)}- 8\vartheta_{(n)}= 0\,,  \label{htheta''}
\end{equation}
with $\lambda_H=1/H$ and $\lambda_n=a/n$ representing the Hubble horizon and the physical scale of the peculiar velocity perturbation respectively. The above solves analytically giving  $\tilde{\vartheta}\propto t^{2/3}\propto a$ on super-Hubble lengths (where $\lambda_H\ll\lambda_n$). Near and inside the horizon, on the other hand, we find that the growth-rate of $\vartheta$ is reduced, with $\tilde{\vartheta}\rightarrow$~constant well inside the Hubble radius. In either case the relativistic growth-rate is significantly stronger than the Newtonian one, where $\tilde{\vartheta}\propto t^{-1/3}\propto a^{-1/2}$ (see solution (\ref{ltvtheta}) in \S~\ref{ssLEPVSS}).

Finally, when dust dominates the energy density of the universe, the homogeneous part of Eq.~(\ref{lGRvsigma''}) takes the form
\begin{equation}
9t^2\tilde{\varsigma}_{ab}^{\prime\prime}+ 15t\tilde{\varsigma}_{ab}^{\prime}- 8\tilde{\varsigma}_{ab}= 0\,. \label{tvsigma''2}
\end{equation}
with an exactly analogous expression monitoring the evolution of the peculiar vorticity. As a result,  after matter-radiation equality and on scales where the inhomogeneous components of (\ref{lGRvsigma''}) and (\ref{lGRvpi''}) are negligible, we obtain $\varsigma$, $\varpi\propto t^{2/3}\propto a$. Again, the relativistic growth-rates are considerably stronger than their Newtonian counterparts (compare to solutions (\ref{ltvpi}) and (\ref{ltvsigma}) in \S~\ref{ssLSPVs} and \S~\ref{ssLEPVSS} respectively).

Before closing this section, we should point out that the relativistic solutions provided here are identical to those reported in~\cite{TT}, where we refer the reader for further discussion. This agreement results from the one between the corresponding differential equations, although they differ in their frame choice. More specifically, the analysis of~\cite{TT} is done in the coordinate system of the matter, namely along the $\tilde{u}_a$-field, whereas ours takes place in the quasi-Newtonian frame (aligned along $u_a$). Practically speaking, the overall agreement of the two approaches reflects the fact that the assumption of an irrotational and shear free $u_a$-field made in~\cite{TT} was never explicitly applied during their derivation. If it were, it would have affected the evolution of the peculiar vorticity (see footnote~13 in \S~\ref{sssEF} previously).

\section{Discussion}\label{sD}
Large scale peculiar velocities, often referred to as bulk flows, appear to be quite common in the universe. This has been established observationally by numerous surveys, although the scale and the velocity of the measured peculiar motions remain under debate (see~\cite{ND,WFH} for representative studies). The theoretical work investigating the evolution of cosmological peculiar-velocity fields has been primarily Newtonian, or quasi-Newtonian, in nature. Also, the majority of the studies focus on the peculiar velocity itself and typically bypass the rest of the kinematics variables, namely the expansion/contraction, the shear and the vorticity of the peculiar motion, with the exception (to the best of our knowledge) of~\cite{TT}. The latter is a fully relativistic linear (perturbative) analysis of the whole spectrum of the peculiar kinematics, with analytical solutions indicating growth-rates considerably stronger than the Newtonian treatments.

Identifying and explaining the differences between the Newtonian and the relativistic results is one of the main aims of our work. To achieve this, we have adopted the Newtonian version of the (relativistic) 1+3~covariant approach to cosmology used in~\cite{M} and also in~\cite{TT}. Our results are in agreement with those of the earlier Newtonian studies and of the quasi-Newtonian treatments of~\cite{M}, but not with those of~\cite{TT}. More specifically, the Newtonian/quasi-Newtonian growth-rates of the peculiar velocity field appear considerably weaker than their relativistic analogues obtained in~\cite{TT}. In order to identify the reasons behind this disagreement, we compared our Newtonian analysis to a relativistic study. Here, however, we followed the quasi-Newtonian treatment of~\cite{M}, rather than that of~\cite{TT}. This enabled us to recover the results of the latter analysis within the quasi-Newtonian framework and also reveal the reason responsible for the aforementioned disagreement. During the process, it also became clear that some of the kinematic assumptions made in~\cite{TT} could be relaxed (see footnote~9 in \S~\ref{sssR4-A} earlier), which in turn makes the relativistic studies more flexible and broadens their range.

Generally speaking, at the ``root of the problem'' lies the different way Newton's and Einstein' theories treat issues as fundamental as the nature of gravity itself (see \S~\ref{ssRPMs} here for a discussion). More specifically, in relativity, the energy flux also contributes to the stress-energy tensor of the matter and therefore to the local gravitational field. When dealing with peculiar flows, such a flux contribution comes from the drift motion of the matter. In a sense, one could say that the peculiar flow itself gravitates~\cite{TT}. This, purely relativistic effect feeds into the Einstein field equations and eventually emerges in the evolution formulae of the peculiar-velocity perturbations. Technically speaking, the effect propagates via the 4-acceleration, which makes the latter the key to the relativistic study of the subject. In the earlier quasi-Newtonian studies the 4-acceleration was expressed in terms of an effective gravitational potential, analogous to the one used in the purely Newtonian treatments. Moreover, the time-evolution of the effective potential (and therefore of the 4-acceleration) followed from an ansatz. Here, as well as in~\cite{TT}, the 4-acceleration was derived analytically from linear cosmological perturbation theory. Therefore, the disagreement between the quasi-Newtonian treatments of~\cite{M} and that of~\cite{TT} is just an apparent one. This does not apply to the purely Newtonian studies, however, since the latter do not allow for any flux contribution to the gravitational field. As a result, Newton's theory cannot reproduce Eq.~(\ref{lGRAa}), which may therefore be seen as the relativistic correction responsible for the disagreement with the Newtonian results and conclusions (with a possible exception in the case of the vorticity -- see footnote~13). This lack of agreement should make one cautious when using Newton's rather than Einstein's theory to study the kinematics of bulk peculiar flows, especially on large (cosmological) scales. The risk is that, by adopting the Newtonian approach, one could seriously underestimate the velocities and the overall kinematic evolution of cosmological peculiar flows. On these grounds, the relativistic analysis seems to support a number of surveys (e.g.~\cite{WFH}) reporting large-scale bulk flows with peculiar velocities faster than it has been typically anticipated.\vspace{2mm}

\textbf{Acknowledgements:} KF acknowledges support from the General Secretariat for Research and Technology (G.S.R.T.) and from the Hellenic Foundation for Research and Innovation (H.F.R.I.), under the “First Call for Research Projects to support Doctoral Candidates” (Project Number: 114). CGT was supported by the Hellenic Foundation for Research and Innovation (H.F.R.I.), under the “First Call for H.F.R.I. Research Projects to support Faculty members and Researchers and the procurement of high-cost research equipment grant” (Project Number: 789).

\end{document}